\documentclass[12pt, letterpaper]{article}
\usepackage{jheppub}
\usepackage{bbm}
\usepackage{color}
\usepackage{amsmath}
\usepackage{amssymb}

\newcommand{\be}{\begin{equation}}
\newcommand{\ee}{\end{equation}}
\newcommand{\ev}[1]{\left\langle#1\right\rangle}
\newcommand{\ket}[1]{\left|#1\right\rangle}
\newcommand{\bra}[1]{\left\langle#1\right|}
\newcommand{\oprod}[2]{\left|#1\right\rangle\!\!\left\langle#2\right |}

\title{Thermal Corrections to Entanglement Entropy from Holography}

\author{Stefan Leichenauer}
\affiliation{Center for Theoretical Physics and Department of Physics,\\
 University of California, Berkeley, CA 94720, U.S.A.}
\emailAdd{sleichen@berkeley.edu}

\abstract{We use holographic techniques to calculate the first thermal correction to the entanglement entropy of a cap-like region of a CFT defined on a sphere, successfully reproducing the field theory result. Since this is an order-one correction to the entropy in the large-$N$ expansion, quantum corrections to the holographic entanglement entropy formula are essential. The bulk calculation is made tractable using the same technical machinery recently used to derive the linearized Einstein equations in the bulk from the first law of entanglement entropy in the CFT.}

\begin{document}
\maketitle

\section{Introduction}

Herzog~\cite{Herzog:2014fra} has recently calculated the first thermal correcton to the entanglement entropy of a cap-like region $B$ for a CFT defined on $R\times S^{d-1}$. The CFT is assumed to have a mass gap $\Delta$ and degeneracy $g$ in the first excited state, in terms of which we have
\be\label{eq-answer}
\delta S^{\rm CFT}_B(T) \equiv S_B(T) - S_B(0) = g\Delta e^{-\beta \Delta}I_d (\theta_0) + o(e^{-\beta \Delta})
\ee
where
\be
I_d(\theta_0) = 2\pi\frac{\Omega_{d-2}}{\Omega_{d-1}}\int_0^{\theta_0}d\theta \frac{\cos\theta - \cos\theta_0}{\sin\theta_0} \sin^{d-2}\theta,
\ee
the opening angle of the cap is $2\theta_0$, $\Omega_{d-1}$ is the volume of $S^{d-1}$, and the CFT is at temperature $T = \beta^{-1}$.\footnote{Herzog argues in Ref.~\cite{Herzog:2014fra} that the expression on the right-hand side of (\ref{eq-answer}) can shift depending on contributions of boundary terms, such as those coming from the conformal coupling of a free scalar to the background curvature. In Ref.~\cite{Casini:2014yca}, the claim is that a generic interacting CFT will not have such boundary terms. We will reproduce (\ref{eq-answer}) as written, and leave to future work the incorporation of boundary terms for non-generic interacting theories.} We are working in units where the sphere has radius one.

The result (\ref{eq-answer}) is independent of $N$ in the sense of large-$N$ CFTs, and so in the large-$N$ expansion should appear at $O(1)$. To calculate $O(1)$ corrections to the entropy holographically, we need to use the quantum generalization of the Ryu-Takayanagi~\cite{Ryu:2006bv} entropy formula due to Faulkner, Lewkowycz, and Maldacena~\cite{Faulkner:2013ana}:
\be
S = \frac{A}{4G_N} + S_{\rm bulk},
\ee
where $A$ is the area of the extremal surface anchored on the boundary of $B$ and $S_{\rm bulk}$ is the entanglement entropy of the bulk matter in the region between $B$ and the extremal surface.\footnote{For general theories of matter coupled to gravity, we would have to use a Wald-like entropy formula in place of the area and the associated terms. While we restrict ourselves to theories where entropy is represented by area for simplicity, the techniques of Ref.~\cite{Faulkner:2013ica} that we employ, based on the formalism of Iyer and Wald~\cite{Iyer:1994ys}, can be used in the more general case.} We will also note that, in finding the difference between the $T=0$ and finite $T$ entropies, there are some fruitful cancelations. From the CFT point of view there are short-distance divergences in the entropy which cancel in the difference; in the bulk this comes from cancelations near the boundary at infinity. But there are similar short-distance divergences in $S_{\rm bulk}$ that also cancel, which we would otherwise have to regulate with counterterms. To summarize, we must compute
\be
\delta S^{\rm holo}_B(T) \equiv \frac{\delta A(T)}{4G_N} + \delta S_{\rm bulk}(T) ,
\ee
and show that it is equal to $\delta S_B^{\rm CFT}$, where the vacuum-subtracted area, $\delta A(T)$, and vacuum-subtracted bulk entropy, $\delta S_{\rm bulk}(T)$, are both finite, well-behaved quantities.

At high temperatures, above the Hawking-Page phase transition, we normally say that the geometry dual to the thermal state is a large black hole. Alternatively, one can say that all of the typical pure states in the canonical ensemble at high temperature look like the same large black hole, and so thermal expectation values can be computed using that black hole background.\footnote{Up to subtleties involving the near-horizon region.} Below the Hawking-Page phase transition, this is not the case. In the low-temperature limit, the thermal state can be expanded as
\be\label{eq-lowTstate}
\rho(T) = \oprod{0}{0} + e^{-\beta\Delta}\left(\oprod{\Delta}{\Delta} - \oprod{0}{0}\right) + \cdots.
\ee
Here we are denoting by $\ket{0}$ the vacuum state (i.e., empty AdS), and $\ket{\Delta}$ the bulk field state with energy $\Delta$ (for notational simplicity we are assuming no degeneracy), which for example could be the lowest-energy single-particle state of a free scalar field of mass $m^2 = \Delta(\Delta -d)$. The bulk represented by this mixed state not a single geometry, but is just what the equation implies: an incoherent mixture of the bulk vacuum and the lowest-lying excited bulk state.\footnote{The ``classical" $O(1)$ part of the geometry is of course the same in both cases, but for our purposes the ``quantum" $O(1/N^2)$ part is also needed.}  We need to compute the extremal surface area in this mixed state, as well as the bulk entanglement entropy, and subtract the respective vacuum values. We will see that both of these computations reduce to calculations in the state $\ket{\Delta}$, which we will perform.
 
\section{Geometric Setup}\label{sec-geometry}

\begin{figure}[t]\hspace*{-.5cm}
\centerline{\includegraphics[width=0.4\columnwidth]{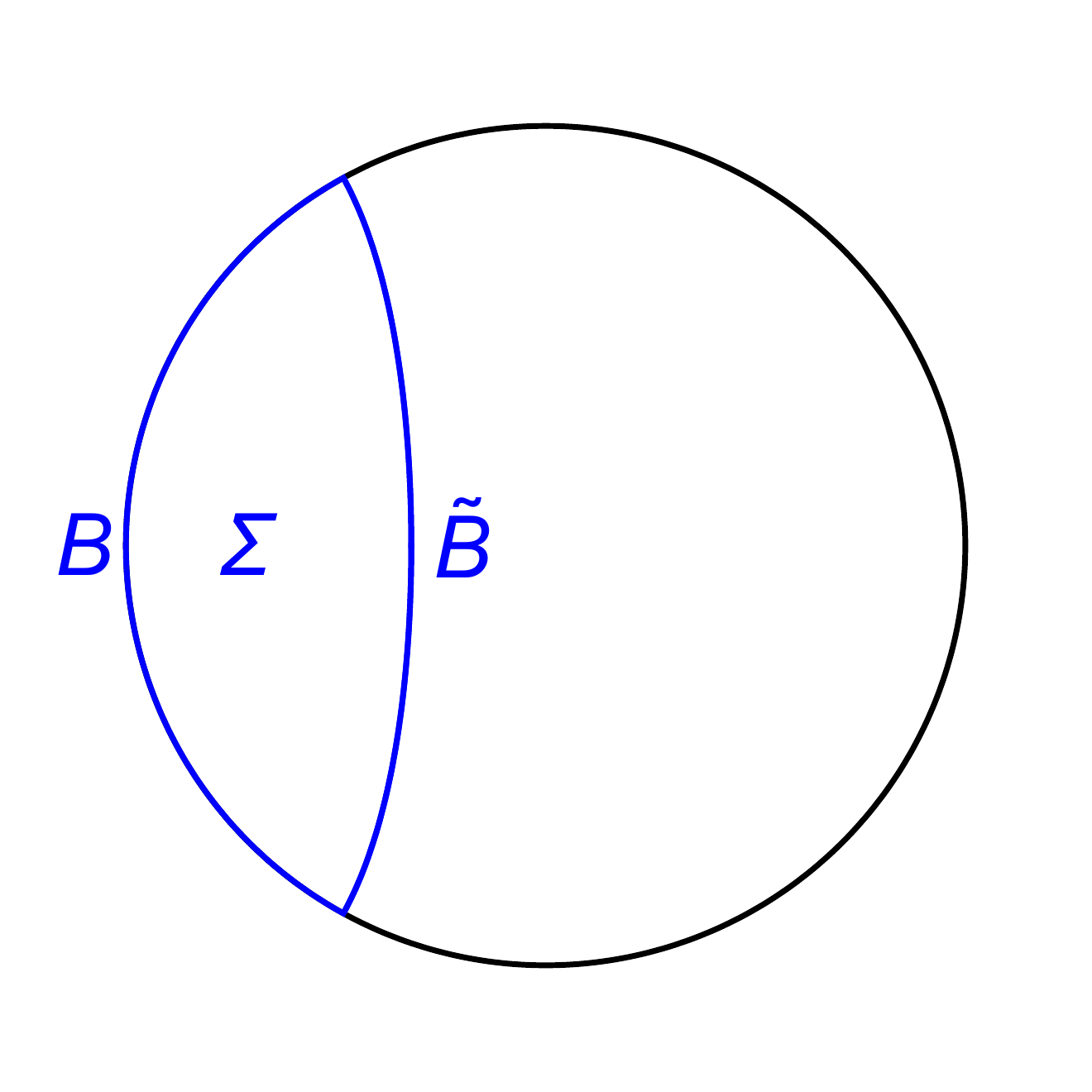},\includegraphics[width=0.4\columnwidth]{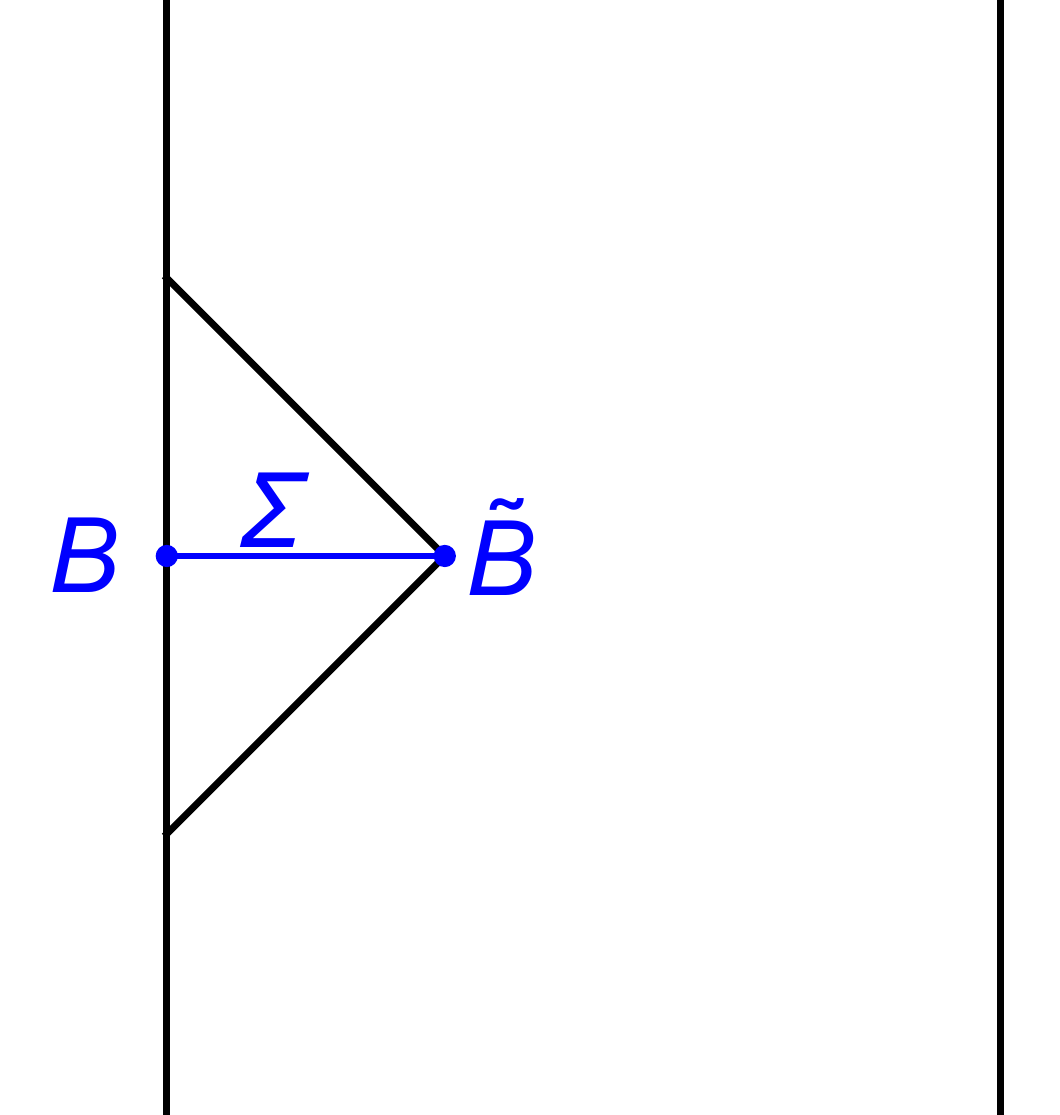}}
\caption{(Left) The $\tau=0$ slice of AdS. The region $B$ is on the boundary, $\tilde{B}$ is the bulk extremal surface, and $\Sigma$ is between them in the bulk. (Right) A cross-section of AdS, where time runs vertically. $\Sigma$ lies at $\tau=0$ and the causal wedge associated to $B$ is shown.}\label{fig}
\end{figure}

The region $B$ on the sphere whose entropy we are computing is defined by its opening angle $2\theta_0$. In terms of the polar angle $\theta$, it is the region $\theta < \theta_0$. In the bulk, we will typically use global coordinates where the metric takes the form
\be
ds^2 = \frac{1}{\cos^2\rho}\left(-d\tau^2 + d\rho^2 + \sin^2\rho\, d\Omega_{d-1}^2 \right),
\ee
where $d\Omega_{d-1}^2$ is the metric of a unit $(d-1)$-sphere, and the angle $\theta$ is one of the coordinates on this sphere:
\be
d\Omega_{d-1}^2 = d\theta^2 + \sin^2\theta d\Omega_{d-2}^2.
\ee
The boundary is located at $\rho = \pi/2$. The extremal surface $\tilde{B}$ which shares its boundary with $B$ and is used to calculate the entropy is given by the equation
\be
\sin \rho \cos \theta = \cos\theta_0.
\ee
The region between $B$ and $\tilde B$ on the $\tau =0$ surface will be denoted by $\Sigma$ (see Fig.~\ref{fig}). When we compute the bulk entanglement entropy, we can think of it as computing the entanglement entropy of the state restricted to $\Sigma$.

It so happens that, since we are considering such a simple surface in empty AdS, $\tilde B$ is also the boundary of the ``causal wedge" associated to $B$. The causal wedge in the bulk is the set of all points whose past and future intersects the domain of dependence of $B$ on the boundary. We can cover the causal wedge with coordinates $t$, $\zeta$, and $u$ defined by
\begin{align}
\cosh \zeta \sinh u &=  \tan\rho \sin\theta,\\
\sinh \zeta \sinh t &=  \frac{\sin \tau}{\cos\rho},\\
\cosh \zeta \cosh u + \cos\theta_0 \sinh \zeta \cosh t &= \sin\theta_0 \frac{\cos \tau}{\cos\rho},\\
\cos\theta_0\cosh \zeta \cosh u + \sinh \zeta \cosh t &= \sin\theta_0  \tan\rho \cos\theta,
\end{align}
in terms of which the metric is of AdS-Rindler form:
\be
ds^2 = -\sinh^2\zeta dt^2 + d\zeta^2 + \cosh^2\zeta(du^2 + \sinh^2 u d\Omega_{d-2}^2).
\ee
We see that on the boundary, $\zeta \to \infty$, this choice of coordinates induces an $R\times H^{d-1}$ geometry on the domain of dependence of $B$. The surface $\tilde B$ is obtained by taking $\zeta\to 0$ at fixed $t$. The Killing vector $\xi^a = \partial_t$ generates time translations in this AdS-Rindler patch, and for calculations it is useful to have an expression for it in terms of the global coordinates:
\be\label{eq-killing}
\xi^a =\frac{\cos\tau\sin\rho\cos\theta - \cos\theta_0}{\sin\theta_0}\partial_\tau +  \frac{ \cos\theta \cos\rho\sin\tau }{\sin\theta_0}\partial_\rho - \frac{ \sin\theta\sin\tau }{\sin\rho\sin\theta_0}\partial_\theta.
\ee
Notice that $\xi^a$ vanishes on $\tilde B$. We have normalized $\xi^a$ so that it has unit surface gravity, $\kappa=1$, where $\xi^a \nabla_a \xi^b = \kappa \xi^b$ on the Rindler horizon.

\section{Holographic Entropy Calculation}

\paragraph{The Area Term}
\hfill\break
First we will compute the area term. According to the Ryu-Takayanagi~\cite{Ryu:2006bv} prescription, we need to find the area of the extremal surface in the bulk which is anchored on the boundary of $B$. As is typically formulated, this term only makes sense when there is a well-defined notion of a classical geometry in the bulk. For our purposes a slight generalization is necessary. We will interpret the Ryu-Takayanagi term in the entropy to be the expectation value of the operator $\hat{A}$ which encodes the area of the extremal surface. This operator is a functional of the metric operator, which for small perturbations we can write as $g+\hat{h}$ with $g$ the AdS vacuum metric. In the low-temperature limit, we use (\ref{eq-lowTstate}) to write the difference between the finite-temperature area and the vacuum area as
\be
\delta A(T) = {\rm Tr}(\hat{A}\rho(T)) -\bra{0}\hat{A}\ket{0} = e^{-\beta \Delta}\left(\bra{\Delta}\hat{A}\ket{\Delta}-\bra{0}\hat{A}\ket{0}\right) +\cdots.
\ee
We are only interested in evaluating this difference to leading order in $G_N$, and so we can expand $\bra{\Delta}\hat{A}\ket{\Delta}$ to first order in $\hat{h}$. Up to operator ordering ambiguities (which do not appear at linear order), $\hat{A}$ should equal $A[g+\hat{h}]$, where $A[\cdot]$ is just the classical functional of the metric which computes the area. Schematically, we have
\be
\bra{\Delta}\hat{A}\ket{\Delta} = \bra{\Delta}A[g+\hat{h}]\ket{\Delta} = A[g] + \left.\frac{\delta A}{\delta h}\right|_{h=0}\bra{\Delta}\hat{h}\ket{\Delta} +\cdots.
\ee
where $A[g]$ is the classical empty AdS value of the area, equal to $\bra{0}\hat{A}\ket{0}$ before quantum corrections.\footnote{As pointed out in Ref.~\cite{Faulkner:2013ana}, quantum corrections to the area term are also important at $O(1)$. But since quantum corrections are already down by a power of $N^2$, the difference in those corrections between the vacuum and first excited states will be further suppressed.} The result is that the first-order corrections to the quantum expectation value for the area can be computed by substituting the expectation value for the metric perturbation into the classical area functional, and expanding to first order. This is not the case for the higher-order corrections if the metric perturbation behaves non-classically, which is as we would expect from a single-particle state with a broad wavefunction like $\ket{\Delta}$.

Furthermore, to first order in $G_N$ the expectation value of the metric perturbation is obtained easily from the expectation value of the linearized Einstein equations:
\be
E_{\mu\nu}[\langle \hat{h} \rangle] = \ev{ E_{\mu\nu}[\hat{h}]} = 8\pi G_N \ev{\hat{T}^{\rm matter}_{\mu\nu}},
\ee
where we have introduced the notation $E_{\mu\nu}$ to denote the linearized equation of motion for the metric perturbation in the absence of matter (but including the cosmological constant).\footnote{$\hat{T}^{\rm matter}_{\mu\nu}$ can still contain contributions from gravitons, and the excited state in question could be a pure graviton state. In that case, the metric perturbation would still have to self-consistently satisfy the Einstein equations as we have written them to this order in $G_N$. For example, the expectation value of the metric at infinity must encode the total energy of the graviton state.} To summarize, at this order in perturbation theory we can obtain the correct backreaction of the matter on the geometry by treating $\langle\hat{T}^{\rm matter}_{\mu\nu}\rangle$ as a classical source. From now on we will simplify our notation by writing $\langle \hat{h}_{\mu\nu} \rangle = h_{\mu\nu}$ and $\langle\hat{T}^{\rm matter}_{\mu\nu}\rangle= T^{\rm matter}_{\mu\nu}$.

Now it is a simple matter to find $\delta A(T)$. The first-order correction to the extremal surface area is obtained by leaving the surface fixed and integrating the linear deviation of the area functional over that surface:
\be\label{eq-deltaA}
\delta A(T)= e^{-\beta \Delta}\int_{\tilde {B}} d^{d-1}x\, \left.\frac{\delta A}{\delta h_{\mu\nu}(x)}\right|_{h=0} h_{\mu\nu}(x).
\ee

\paragraph{Bulk Entropy}
\hfill\break
The calculation of the bulk entropy is similar to the CFT calculation in Ref.~\cite{Herzog:2014fra}. In the low-temperature limit, the reduced density operator for the region $\Sigma$ is found by taking the trace of (\ref{eq-lowTstate}) over the complement of $\Sigma$:
\be
\rho_\Sigma(T) = \rho_\Sigma^0 + e^{-\beta\Delta}( \rho_\Sigma^\Delta - \rho_\Sigma^0) + \cdots
\ee
The second term is just a small perturbation, so we can apply the first law of entanglement entropy:
\be\label{eq-modular}
\delta S_{\rm bulk}(T) = -e^{-\beta\Delta}{\rm Tr}\left[\left( \rho_\Sigma^\Delta - \rho_\Sigma^0\right)\log \rho_\Sigma^0\right]= e^{-\beta\Delta}\delta K_\Sigma,
\ee
where we have denoted by $\delta K_\Sigma$ the change in the expectation value of the vacuum modular Hamiltonian on $\Sigma$, $-\log \rho_\Sigma^0$. Recall that $\Sigma$ is a constant-time surface in the AdS-Rindler space constructed in Sec.~\ref{sec-geometry}. Much like flat Rindler space, the vacuum state of AdS-Rindler is obtained by a $2\pi$ tranlsation in Euclidean AdS-Rindler time. So the vacuum modular Hamiltonian in AdS-Rindler space is just $2\pi$ times time-translation Hamiltonian associated to that space, i.e., translation by $\xi^a$, which is the integral over $\Sigma$ of one of the components of the bulk stress tensor. Thus we have
\be\label{eq-deltaS}
\delta S_{\rm bulk}(T) = 2\pi e^{-\beta\Delta}\int_\Sigma \xi^\mu T^{\rm matter}_{\mu \nu}d\Sigma^\nu,
\ee
where $\xi^a = \partial_t$ is the AdS-Rindler Killing vector introduced previously, and $d\Sigma^a$ is the volume form on $\Sigma$.

\paragraph{Putting Things Together}
\hfill\break
Now we just have to put together (\ref{eq-deltaA}) and (\ref{eq-deltaS}) to get
\begin{align}
\delta S^{\rm holo}_B(T) &\equiv \frac{\delta A(T)}{4G_N} + \delta S_{\rm bulk}(T)\\
&= e^{-\beta \Delta} \left[ \frac{1}{4G_N}\int_{\tilde {B}} d^{d-1}x \frac{\delta A}{\delta h_{\mu\nu}(x)} h_{\mu\nu}(x) +2\pi\int_\Sigma \xi^\mu T^{\rm matter}_{\mu \nu}{\bf \epsilon}^\nu \right].
\end{align}
To see that this is equivalent to $\delta S^{\rm CFT}_B$ in (\ref{eq-answer}), we make use of the $(d-1)$-form ${\bf \chi}$ constructed out of the metric perturbation $h$ in Ref.~\cite{Faulkner:2013ica}, which has the (off-shell) properties
\begin{align}
\int_{\tilde B} {\bf \chi} &= \frac{1}{4G_N}\int_{\tilde {B}} d^{d-1}x \left.\frac{\delta A}{\delta h_{\mu\nu}(x)}\right|_{h=0} h_{\mu\nu}(x),\\
\left.d{\bf \chi}\right|_\Sigma &= -\frac{1}{4 G_N} \xi^\mu E_{\mu\nu}[h] d\Sigma^\nu.
\end{align}
Additionally, the integral of $\chi$ over the surface $B$, which from the bulk point of view is the part of the boundary of $\Sigma$ located at infinity, can be written as
\be\label{eq-chiB}
\int_{B} {\bf \chi} = \int_B d^{d-1}x \,\delta S_B^{\rm grav}[h],
\ee
where we still need to identify $\delta S_B^{\rm grav}[h]$. To do so, recall the holographic dictionary for the metric perturbation near the boundary (schematically):
\be\label{eq-dict}
\lim_{\rho \to \pi/2} \left(\frac{\pi}{2}-\rho\right)^{2-d}h_{\mu\nu}  \sim \delta \ev{T_{\mu \nu}^{\rm CFT}},
\ee
where $\delta \ev{T_{\mu \nu}^{\rm CFT}}$ is the vacuum-subtracted expectation value of the stress tensor in the CFT. The region $B$ is simple enough that $\delta K_B$, the vacuum-subtracted expectation value of the CFT vacuum modular Hamiltonian on $B$, can be written as a local integral of $\delta \ev{T_{\mu \nu}^{\rm CFT}}$:
\be
\delta K_B \equiv -{\rm Tr}\left[\left( \rho_B - \rho_B^0\right)\log \rho_B^0\right] = \int_B \left(\text{local function of }\delta \ev{T_{\mu \nu}^{\rm CFT}}\right).
\ee
So, applying (\ref{eq-dict}), $\delta K_B$ can be written as a local integral of the asymptotic value of the metric perturbation $h$ in the bulk. That local integral is precisely the right-hand side of (\ref{eq-chiB}), which defines $\delta S_B^{\rm grav}$. Said another way, the integral of $\chi$ over $B$ is $\delta K_B$ in the CFT translated into bulk language using the holographic dictionary (\ref{eq-dict}).

If we evalutate (\ref{eq-chiB}) in the state $\ket{\Delta}$ we will find $\delta K_B$ in the state $\ket{\Delta}$. By including the Boltzmann factor, we obtain the first thermal correction to the CFT entanglement entropy of $B$ in a way entirely parallel with the bulk calculation in (\ref{eq-modular}), and in fact this is how (\ref{eq-answer}) was obtained by Herzog~\cite{Herzog:2014fra}. In other words, we have
\be
e^{-\beta \Delta} \int_{B} {\bf \chi} =e^{-\beta \Delta} \int_B dx^{d-1} \delta S_B^{\rm grav}[h] = \delta S^{\rm CFT}_B(T),
\ee
where $\delta S^{\rm CFT}_B(T)$ is the quantity appearing in (\ref{eq-answer}). We would like to emphasize that this follows directly from the definition of $\delta S_B^{\rm grav}$ and the property (\ref{eq-chiB}) of $\chi$.

Applying Stokes' theorem to $\chi$ yields\footnote{The sign coventions are such that the boundary of $\Sigma$ is $\partial \Sigma = \tilde{B} - B$.}
\be
-\int_\Sigma \frac{1}{4 G_N} \xi^\mu E_{\mu\nu}[h] d\Sigma^\nu = \int_\Sigma d\chi = \int_{\tilde B} \chi -\int_B \chi = e^{\beta \Delta}\left[\frac{\delta A(T)}{4G_N}- \delta S^{\rm CFT}_B(T)\right]~.
\ee
The linearized Enstein equations, $E_{\mu\nu} = 8\pi G_N T_{\mu\nu}^{\rm matter}$, then give
\be
\delta S^{\rm CFT}_B(T) = \frac{\delta A(T)}{4G_N} + 2\pi e^{-\beta \Delta} \int_\Sigma \xi^\mu T_{\mu\nu}^{\rm matter}d\Sigma^\nu  = \frac{\delta A(T)}{4G_N}+ \delta S_{\rm bulk}(T) =  \delta S^{\rm holo}_B(T),
\ee
completing the argument. Note that this guarantees that $\delta S^{\rm CFT}_B = \delta S^{\rm holo}_B$ even without knowing the explicit form of $\delta S^{\rm CFT}_B(T)$. As an illustration, we go through an example in Appendix~\ref{sec-example} where the exact expression in (\ref{eq-answer}) is reproduced directly from the bulk.

\section{Discussion}

We have shown that the first thermal corrections to CFT entanglement entropy can be produced from a purely bulk calculation. There are two main ideas that came together to accomplish this. First, as was already noticed in Refs.~\cite{Faulkner:2013ica, Swingle:2014uza}, there is a deep connection between the linearized Einstein equations in the bulk, the modular Hamiltonian on the boundary, and the linearized area functional on the extremal surface. The context here is a little different, though. In Refs.~\cite{Faulkner:2013ica, Swingle:2014uza}, the bulk geometry was assumed classical, and the differential form $\chi$ was used to derive the linearized Einstein equations for that classical metric from the properties of entanglement entropy on the boundary. Here we are {\em assuming} the Einstein equations, in the form of operator equations inside of expectaion values, and using them to learn something about the entanglement entropy. The differential form $\chi$ connects the bulk and the boundary in the same way, but the direction of the argument has switched around.

The second main idea was in our treatment of the Ryu-Takayanagi area functional. The metric backreaction was not assumed classical, but still gave a contribution to the area of the extremal surface. The idea that the Ryu-Takayanagi term should be thought of as an expectation value is very natural, but is difficult to test in a nontrivial way. Here we really only used the mild assumption that it behaves as an expectation value with respect to thermal averaging and linear quantum corrections. For higher-order corrections, it would be important to treat this term in the correct way (which likely involves using the generalized entropy in the bulk instead of the area and bulk entanglement separately~\cite{Engelhardt:2014gca}). It would be very interesting if a more nontrivial example could be found, such as one where off-diagonal matrix elements of $\hat{A}$ were important.

\acknowledgments

I would like to thank Chris Akers, Raphael Bousso, Zach Fisher, Chris Herzog, Jason Koeller, and Mudassir Moosa for useful discussions and correspondence. This work was supported in part by the Berkeley Center for Theretical Physics, by the National Science Foundation (award numbers 1214644 and 1316783), by the Foundational Questions Institute grant FQXi-RFP3-1323, by ÒNew Frontiers in Astronomy and CosmologyÓ, and by the U.S. Department of Energy under Contract DE-AC02-05CH11231.

\appendix

\section{Example: Bulk Scalar Field}\label{sec-example}

In this appendix we will work through an example where the lowest-energy state is that of a free scalar field in the bulk, of mass $m^2 = \Delta(\Delta -d)$. The lowest energy state is the one with zero angular momentum and zero radial quantum number. Let $a^\dagger$, $a$ be the creation and annihilation operators for this state, in terms of which the field can be written as
\be
\phi = N (a e^{i\Delta \tau} + a^\dagger e^{-i\Delta \tau})\cos^\Delta \rho + \cdots,
\ee
where $N$ is a normalization constant to be fixed later and the $\cdots$ represent creation/annihilation operators for other modes which do not concern us. In the state $\ket{\Delta} = a^\dagger \ket{0}$, the the expectation value of the energy-momentum tensor is spherically symmetric and time-independent, and in particular the energy density is
\be
T^{\rm matter}_{\tau\tau} = N^2 \Delta\left(\Delta - \frac{d}{2}\right)\cos^{2\Delta - 2}\rho.
\ee
The coefficient $N^2$ can be computed by demanding that the total energy is equal to $\Delta$. In fact, all we need to know about the energy-momentum tensor is that it is time-independent, spherically symmetric, and that the total energy is $\Delta$. Because of the symmetries, the perturbed metric looks like a black hole with a radially-varying mass. It is most convenient to add the perturbation to the global metric using the coordinate $r = \tan \rho$, in which case we have to first order
\be
ds^2 = -\left(1+ r^2 - \frac{\mu(r)}{r^{d-2}}\right)d\tau^2 + \frac{1}{1+r^2}\left(1 + \frac{\mu(r)}{(1+r^2)r^{d-2}}\right)dr^2 + r^2 d\Omega_{d-1}^2,
\ee
with
\be\label{eq-mu}
\mu(r) = \frac{16\pi G_N}{d-1} \int_0^r dr'\frac{r'^{d-1}}{1+r'^2} \,T^{\rm matter}_{\tau\tau}(r').
\ee
It's straightforward to check that the deviation in the area of $\tilde{B}$ due to this metric perturbation is
\be
\delta A(T) =  \frac{e^{-\beta \Delta}\Omega_{d-2} }{2}\int_0^{\theta_0} d\theta~\sin^{d-2}\theta \frac{\tan^2\theta}{\tan\theta_0 }  \mu(\rho(\theta)),
\ee
where we use $\sin\rho\cos\theta = \cos\theta_0$ for the surface $\tilde B$. By substituting the expression (\ref{eq-killing}) for $\xi^a$ into equation (\ref{eq-deltaS}) for the bulk entropy we find
\be
\delta S_{\rm bulk}(T) = 2\pi  \Omega_{d-2}e^{-\beta \Delta}  \int_\Sigma d\rho d\theta \sin^{d-2}\theta \tan^{d-1}\rho\frac{\sin\rho\cos\theta - \cos\theta_0}{\sin\theta_0} \,T^{\rm matter}_{\tau\tau}(\rho).
\ee
We know from the general discussion of the text that we should make use of Einstein's equation to find an expression for $\mu$ in terms of $T^{\rm matter}_{\tau\tau}$. In this case, the $\tau\tau$-component of Einstein's equation says
\be
T^{\rm matter}_{\tau\tau}(r) = \frac{d-1}{16\pi G_N}\frac{1+r^2}{r^{d-1}}\frac{d\mu}{dr},
\ee
as is evident from (\ref{eq-mu}). Substituting this in for $T^{\rm matter}_{\tau\tau}$ gives
\be
\delta S_{\rm bulk}(T) = \frac{d-1}{8 G_N} \Omega_{d-2}e^{-\beta \Delta}\int_\Sigma d\rho d\theta \sin^{d-2}\theta \frac{\sin\rho\cos\theta - \cos\theta_0}{\sin\theta_0} \frac{d\mu}{d\rho}.
\ee
Being clever, we can choose to rewrite the integrand as
\be
\sin^{d-2}\theta \frac{\sin\rho\cos\theta - \cos\theta_0}{\sin\theta_0} \frac{d\mu}{d\rho} =  \partial_\rho \chi^\theta - \partial_\theta\chi^\rho,
\ee
with
\begin{align}
\chi^\rho &= \frac{1}{d-1}\frac{\cos\rho\sin^{d-1}\theta}{\sin\theta_0}\mu(\rho),\\
\chi^\theta &=\frac{\sin\rho\cos\theta - \cos\theta_0}{\sin\theta_0}\mu(\rho)\sin^{d-2}\theta.
\end{align}
This allows us to use Green's theorem:
\be
\delta S_{\rm bulk}(T) = \frac{d-1}{8 G_N} \Omega_{d-2}e^{-\beta \Delta}\left[\int_0^{\theta_0} d\theta\, \left.\chi^\theta\right|_{\rho = \pi/2} - \int_0^{\theta_0} d\theta\,\left.\left(\chi^\theta+ \frac{d\rho}{d\theta} \chi^\rho\right)\right|_{\sin\rho = \cos\theta_0/\cos\theta}\right].
\ee
The first term reproduces the function in (\ref{eq-answer}),
\be
\frac{d-1}{8 G_N} \Omega_{d-2}e^{-\beta \Delta}\int_0^{\theta_0} d\theta\, \left.\chi^\theta\right|_{\rho = \pi/2}= \Delta e^{-\beta\Delta}I_d(\theta_0) ,
\ee
while the second term gives us the change in area,
\be
 \frac{d-1}{8 G_N} \Omega_{d-2}e^{-\beta \Delta}\int_0^{\theta_0} d\theta\,\left.\left(\chi^\theta+ \frac{d\rho}{d\theta} \chi^\rho\right)\right|_{\sin\rho = \cos\theta_0/\cos\theta} =  \frac{\delta A(T)}{4 G_N}.
\ee
So we have successfully derived (\ref{eq-answer}) from a bulk calculation.

\bibliographystyle{utcaps}
\bibliography{ThermEnt}

\providecommand{\href}[2]{#2}\begingroup\raggedright\begin{thebibliography}{1}

\bibitem{Herzog:2014fra}
C.~P. Herzog, ``{Universal Thermal Corrections to Entanglement Entropy for
  Conformal Field Theories on Spheres},''
  \href{http://dx.doi.org/10.1007/JHEP10(2014)028}{{\em JHEP} {\bf 1410} (2014)
   28},
\href{http://arxiv.org/abs/1407.1358}{{\tt arXiv:1407.1358 [hep-th]}}.

\bibitem{Casini:2014yca}
H.~Casini, F.~Mazzitelli, and E.~T. Lino, ``{Area terms in entanglement
  entropy},''
\href{http://arxiv.org/abs/1412.6522}{{\tt arXiv:1412.6522 [hep-th]}}.

\bibitem{Ryu:2006bv}
S.~Ryu and T.~Takayanagi, ``{Holographic derivation of entanglement entropy
  from AdS/CFT},'' \href{http://dx.doi.org/10.1103/PhysRevLett.96.181602}{{\em
  Phys.Rev.Lett.} {\bf 96} (2006)  181602},
\href{http://arxiv.org/abs/hep-th/0603001}{{\tt arXiv:hep-th/0603001
  [hep-th]}}.

\bibitem{Faulkner:2013ana}
T.~Faulkner, A.~Lewkowycz, and J.~Maldacena, ``{Quantum corrections to
  holographic entanglement entropy},''
  \href{http://dx.doi.org/10.1007/JHEP11(2013)074}{{\em JHEP} {\bf 1311} (2013)
   074},
\href{http://arxiv.org/abs/1307.2892}{{\tt arXiv:1307.2892}}.

\bibitem{Faulkner:2013ica}
T.~Faulkner, M.~Guica, T.~Hartman, R.~C. Myers, and M.~Van~Raamsdonk,
  ``{Gravitation from Entanglement in Holographic CFTs},''
  \href{http://dx.doi.org/10.1007/JHEP03(2014)051}{{\em JHEP} {\bf 1403} (2014)
   051},
\href{http://arxiv.org/abs/1312.7856}{{\tt arXiv:1312.7856 [hep-th]}}.

\bibitem{Iyer:1994ys}
V.~Iyer and R.~M. Wald, ``{Some properties of Noether charge and a proposal for
  dynamical black hole entropy},''
  \href{http://dx.doi.org/10.1103/PhysRevD.50.846}{{\em Phys.Rev.} {\bf D50}
  (1994)  846--864},
\href{http://arxiv.org/abs/gr-qc/9403028}{{\tt arXiv:gr-qc/9403028 [gr-qc]}}.

\bibitem{Swingle:2014uza}
B.~Swingle and M.~Van~Raamsdonk, ``{Universality of Gravity from
  Entanglement},''
\href{http://arxiv.org/abs/1405.2933}{{\tt arXiv:1405.2933 [hep-th]}}.

\bibitem{Engelhardt:2014gca}
N.~Engelhardt and A.~C. Wall, ``{Quantum Extremal Surfaces: Holographic
  Entanglement Entropy beyond the Classical Regime},''
  \href{http://dx.doi.org/10.1007/JHEP01(2015)073}{{\em JHEP} {\bf 1501} (2015)
   073},
\href{http://arxiv.org/abs/1408.3203}{{\tt arXiv:1408.3203 [hep-th]}}.

\end{thebibliography}\endgroup

\end{document}